\documentclass[12pt]{iopart}					
\usepackage{graphicx}						
\usepackage{amssymb}

\begin{document}
\title{On the Anomalous Weight Losses of High Voltage Symmetrical Capacitors}

\author{Elio B. Porcelli$^1$ and Victo S. Filho$^1$}

\address{$^1$H4D Scientific Research Laboratory, 04674225,     S\~ao Paulo, SP, Brazil} 
\ead{vsf7@yahoo.com}

\begin{abstract}
In this work, we analyzed an anomalous effect verified from symmetrical capacitor devices, working in very high electric potentials. The mastery of that effect could mean in the future the possible substitution of propulsion technology based on fuels by single electrical propulsion systems. From experimental measurements, we detected small variations of the device inertia that cannot be associated with known interactions, so that the raised force apparently has not been completely elucidated by current theories. We measured such variations within an accurate range and we proposed that the experimental results can be explained by relations like Clausius-Mossotti one, in order to quantify the dipole forces that appear in the devices. The values of the weight losses in the capacitors were calculated by means of the theoretical proposal and indicated good agreement with our experimental measurements for 7kV and with many other experimental works. \\

\pacs{06.30.Gv,77.22.Jp,84.32.Tt}

\end{abstract}

\maketitle

\section{Introduction}
\label{intro}

In an earlier experimental work~\cite{Manitoba}, it was described that an anomaly concerning to inertia measurements occurs in experiments involving parallel plate capacitor devices when they were subjected to high voltage. This novel effect was named in some works Biefeld-Brown effect, a phenomenon known but not entirely understood since 1928~\cite{Brown1,Brown2,Williams}. Basically, the effect is characterized by 
a small upward force that raises when parallel plate capacitors are subjected to high-voltage. 

In Ref.~\cite{Brown1,Brown2}, Brown developed a basic principle for inducing motion based on the observed tendency of  a capacitor charged by high voltage to exhibit motion towards one of its poles. 
The experiments were performed with very high voltages applied on the capacitors, indicating as possible explanations the reduction on their weight or the presence of an anomalous net force acting on the capacitors~\cite{Manitoba,Brown1,Brown2}. Despite of the weakness of the possible interaction, the observed force consistently moved the charged capacitor in an upward direction when their poles were aligned in the vertical position. 

In a first moment, one could imagine that the responsible agent for the effect was the electromagnetic force. But, such a detected force could not be associated with the electromagnetic interaction due to fields of the Earth because it was shown that the upward motion was proportional to the magnitude of the electrical potential energy stored in the electrostatic field of the capacitor. 

Based on this motivation, we investigated in a systematical way the effect, so that we could experimentally verify and quantify that force and propose a possible new theoretical explanation. 
In order to verify that anomalous effect, we implement a series of experimental measurements for detecting its existence in a consistent way. We have adopted an alternative experimental setup, in which it is possible to measure the magnitude of the force with no doubts concerning to its nature.  

In the following, we describe in a more detailed way the experiments performed in some earlier works. In the following section, we detail our experimental setup, as well our measurements concerning to weight variations of the capacitor devices. Afterwards, we describe our theoretical model in order to present an alternative explanation of the phenomenon. Finally, in the last section, we present our conclusions and final remarks. 
\section{The Biefeld-Brown Effect} 
\label{sec:1}

In Ref.~\cite{Manitoba}, it was reported that some experiments were performed in parallel plate charged capacitors and it was verified that an external unknown force moved the entire capacitor in an upward direction. 
Some large symmetrical capacitors used in the experiment were subjected to voltages up to 125 kV, but the materials (a type of glass) presented dielectric breakdown limits, so that the maximum voltage used in order to detect the effect  was 100 kV. It was consistently observed a reduction of their apparent weight even changing the polarity. The Law of Coulomb could not theoretically explain the effect, since that the internal net force should be zero and the force between the Earth and the capacitors did not have the right upward direction. Besides, its magnitude was also several orders of magnitude higher than that one predicted by the traditional electromagnetic theory. Additionally, the total force was significant but weak (not enough to cause the complete levitation of the capacitor) and depended on the magnitude of the electrical potential energy stored on the capacitor. An important point was that its presence was just detected by the upward motion. Those measurements were consistent with the so-called Biefeld-Brown effect. Until the moment, there is no theory that can explain the phenomenon. 

Some other experiments with high voltages applied on capacitors have been realized in 
Refs.~\cite{Mahood,Honda,Woodward1,Woodward2,Woodward3,Poher,Modanese} and they also pointed out the existence of the anomalous forces on those devices. Some effects can not be associated with usual electrostatic forces. 

In Ref.~\cite{Antano}, it was reported that trials to explain the phenomenon were realized, mainly by means of the hypothesis of the existence of an electric wind~\cite{Moller,Tajmar}. However, measurements performed in the vacuum showed that the effect persisted even so~\cite{Brown2,Antano}. Besides, other experiments described in Refs.~\cite{Manitoba,Honda,Musha0} showed that insulating materials around the capacitor devices can not reduce neither eliminate the effect.   

In Ref.~\cite{Mahood}, it has also mentioned that a lot of measurements of inertia in capacitors operating in high voltage have supported the existence of its transient reduction. For instance, in the experimental works described in Refs.~\cite{Woodward1,Woodward2,Woodward3}, quick oscillations of the energy density of high-energy capacitors indicated reduction in weight of up to 10\%. Even so, the effect is still a bit controversial, mainly due to the weakness of the force generated in the experimental process. Besides, it has been suggested that a force arises from Special Relativity Theory, as consequence of Mach's Principle. As known, Mach's Principle tell us that the inertia of an body exists due to the gravitational interaction of it with all the matter in the Universe.  

Despite of that controversy and disagreement among researchers, the high number of works reporting the phenomenon, even with weak  effects, suggests and claims for more investigation, accurated experimental verification and a good theoretical explanation. In addition, it is also interesting to say that the effect has been considered for practical purposes in order to allow the substitution of technologies of propulsion based on fuels by single electrical propulsion systems~\cite{Mahood}.   

The laboratory Honda has also analyzed the phenomenon. In Refs.~\cite{Honda,Musha0}, it was reported that a lot of experiments were performed to verify the Biefeld-Brown effect. The experimental setups were composed by an rounded symmetrical parallel plates capacitor devices in order to verify one of main explanations for the phenomenon, that is, the existence of corona discharges and electric wind around the capacitor. So, they applied AC and DC voltages up to 18 kV on capacitors of materials such as glass and plastic with different thicknesses, some of them enclosed by an insulator oil contained within a metallic vessel. Then, measurements were performed and were found weight losses  
$\Delta$m = ( 0.9 $\pm$ 0.63 )~g. 
However, the forces generated by the ionic transfer of momentum, feature needed for the electric wind hypothesis, can be evaluated by~\cite{Honda,Fazi}:
\begin{equation} 
\Delta M = \frac{i}{g}\sqrt{\frac{2mV}{q}} \ , 
\label{wind}
\end{equation} 
in which $\Delta M$ is the inertia reduction, m the mass of the capacitor, V the applied potential, q the charge on the capacitor, g the gravity acceleration and i the current applied. So, by means of the experimental values in the work, 
it was obtained that the weight reduction due to ionic winds amounts to $\Delta$m=(1.3x10$^{-3})$g, that is, totally negligible if compared with the experimental result associated to the inertia variation. 

In Refs.~\cite{Honda,Musha0}, theoretical treatments were considered in order to explain the effect. In Ref.~\cite{Musha0}, it was considered the hypothesis that charged particles under intense electric field generates a new gravitational field around itself, from which one shows that the weight reduction of a capacitor is proportional to the applied electric field. However, this linear relationship did not correspond to the experimental data measured for all the magnitudes of the electric field. In this work, it was considered the zero-point vacuum field theory (ZPFT), by attributing the effect to the interaction between the vacuum fluctuations and the high potential electric field provided by the capacitor. The value of weight loss calculated from ZPFT amounts to $\Delta M=$1.28 g. Despite of the relative agreement with the average value of the weight losses, such a result shows that the theoretical results can be improved. In fact, the relative error found on the average value was around 0.42, so that there is margin to a better theoretical explanation of the Biefeld-Brown effect. Besides, the error bar is large, hence better experimental effort must also be implemented.  

With that motivation, we here describe a new experimental work in order to measure possible weight reductions in symmetrical capacitors and consider other possible alternative scenario in order to explain the phenomenon, that we describe in the following. In the next section, we describe our experimental work and compare our measurements with the experimental results from other authors~\cite{Manitoba,Mahood,Honda,Poher,Musha0}. Afterwards, in the following section, we explain our theoretical framework and compared our calculations with the experimental results. Finally, in the last section we discuss our main conclusions. 
\section{Experimental Results}
\label{sec:2}

We initially performed some experimental measurements with the objective of verifying the maximum weight loss of a high-potential symmetrical capacitor sample. In our work, we consider a capacitor sample of two parallel rounded plates made by aluminium, with 118.3 pF and 8 cm of diameter, enclosing a plastic dielectric (polystyrene) 1mm tick, with mass 41.154 g and relative permittivity 2.7. 

\begin{figure}[h]
\begin{center}
\includegraphics[width=8cm]{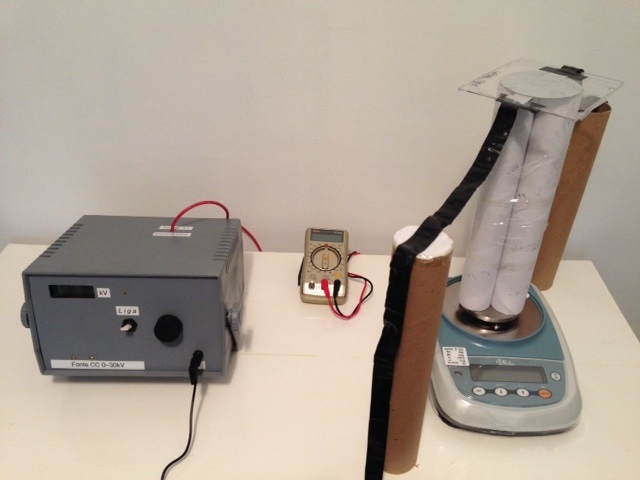}
\caption[dummy0]{Scheme of the experimental setup used for the measurements of the weight losses of the symmetrical capacitors. The rounded parallel plates symmetrical capacitor is indicated at the right side of the picture, connected to the high-voltage power supply. 
The capacitor was upheld on top of a cardboard support placed on the electronic scale, so that its plates were parallel to the scale tray. The electrical connection of the plates of the capacitor with the power supply was made by insulated aluminium stripes, so that one could avoid electrostatic interactions with the scale or with the ambient. The negative output of the power supply and the scale were connected together to the ground of the laboratory.} 
\label{fig1}
\end{center}
\end{figure}

A maximum average weight loss of up to 220 mg was measured when it was applied a maximum DC voltage of 20 kV. Despite of fluctuations on the measurements, it was notable that there is really the existence of an upward force on the capacitor.  
The measurement of the capacitor weight was made by a milligram electronic scale of 300g maximum load.   
The capacitor placed in the horizontal position was supported over the electronic scale by a cylindrical cardboard support of 30 cm height in order to minimize some possible residual electrical interaction between them, according to the scheme shown in Fig.~\ref{fig1}.  
In order to compare, we show in the table \ref{tab1} the main features concerning to the capacitors of the experiments earlier described. 
\begin{table}[h]
{\small{
\caption{Comparison among the physical characteristics or quantities of the parallel plates symmetrical capacitors 
considered in Refs.~\cite{Manitoba,Mahood,Honda,Poher,Musha0} and the correspondent quantities in our experimental work, in order to measure the weight loss or anomalous force of the devices. In the first column, the number before the references represents each different experimental setup used by the corresponding authors.}
\begin{tabular}{|c|c|c|c|c|c|}
\hline
Experiment &	Plate 	& Dielectric	& Dielectric  &	Relative & Voltage\\
Setup & Area   & Tickness & Material & Permittivity  & \ \ \ \\ 
\hline 
1~\cite{Manitoba}	&0.06 m$^2$	&2 mm	              &Glass	            &6	    &100kVDC        	\\
1~\cite{Manitoba}	&0.06 m$^2$	&2 mm	              &Glass	            &6	    &20kVDC         	\\
2~\cite{Manitoba}	&0.03 m$^2$	&2 mm	              &Glass            	&6	    &100kVDC	        \\
2~\cite{Manitoba}	&0.03 m$^2$	&2 mm	              &Glass	            &6	    &20kVDC	         \\
3~\cite{Manitoba}	&3.24 m$^2$	&6.35cm           	&Wax	              &2	    &25kVDC	          \\
4~\cite{Honda}	&227 cm$^2$	&1mm	              &Glass	            &8.37	  & 1 - 8 kVAC	  \\
5~\cite{Poher}	&50 cm$^2$	& \ 0.54 $\mu$m 	& YBa$_2$Cu$_3$O$_5$	    \    &200	  &2863 VAC	       \\
6~\cite{Poher}	&1 cm$^2$	  & \ 0.54 $\mu$m   	& YBa$_2$Cu$_3$O$_5$    \    &200	  &2863 VAC	     \\
7~\cite{Poher}	&25 cm$^2$	& \ 0.54 $\mu$m 	& YBa$_2$Cu$_3$O$_5$	      \  &200	  &2863 VAC	     \\
8~\cite{Mahood}	&0.8 cm$^2$	&2mm	&special mat	&283	              &600VAC	\\
9~\cite{Musha0}	&33.18 cm$^2$	&0.2 mm	&Plastic	&2.3	&1200 VDC\\
9~\cite{Musha0}	&33.18 cm$^2$	&0.2mm	&Plastic	&2.3	&300VDC\\
10~\cite{Musha0}	&227 cm$^2$	&1mm	&Glass	&10	&18kVDC\\
11	&50.26 cm$^2$	&1mm	&Plastic	&2.65	&20kVDC\\
\hline
\end{tabular}
\label{tab1}
}}
\end{table}  

We show in the table \ref{tab1} the physical quantities of our capacitor sample and the correspondent physical quantities of the capacitors used by other authors. All the experiments consider symmetrical capacitors with parallel plates. In that table, the comparison among the data, mainly for the relative permittivity, deserves some comments, as follow. In Ref.~\cite{Manitoba}, it was reported a relative permittivity $\epsilon_r$=6, but just like an estimated value. However, in general we know that as higher the relative permittivity as smaller is the breakdown voltage and vice-versa. In fact, it is known that only pyrex glasses possess a so high breakdown voltage (up to 60 kV/mm), but such special glasses only present $\epsilon_r$ = 3.8, that is, smaller than the estimated one in the article. Even so, we used the value reported in Ref.~\cite{Manitoba} in our calculations, although it is worth to report that the theoretical result is more precise when considering the actual value. 

In Ref.~\cite{Honda}, the relative permittivity mentioned was $\epsilon_r$=10, but if we calculate such a value by using the capacitance informed in the work we obtain $\epsilon_r$ = 8.37. The latter value was used in our calculations. 

In Ref.~\cite{Poher}, it is reported that capacitors made by a superconducting material (YBa$_2$Cu$_3$O$_5$) were used. In this case, the value of the relative permittivity $\epsilon_r$ was calculated by means of the curve of relative permittivity versus temperature of the superconducting material, in the temperature 77 K~\cite{Testardi}. Besides, in the work was only indicated the value of the force (7800 N) for the emitter with 50 cm$^2$; for the other emitters in the table (with diameters 1 cm$^2$ and 25 cm$^2$), the force was calculated according to the information that it is proportional to the emitter area. The anomalous effect just occurs below the critical temperature T$_c$ = 90 K, because the dielectric distance becomes 0.54 $\mu$m. Above the critical temperature, the distance increases to 61 $\mu$m and such a transition enormously decrease the value of the capacitance, so that the effect is unperceivable. We add this material in the table \ref{tab1} because here we are considering that the emitters are like plates of a capacitor, a feature that can be identified directly from Fig.~6 in 
Ref.~\cite{Modanese}. In fact, the material composed by an insulator and a superconducting layer used in the experiment presented the feature of accumulating negative charges at the negative electrode. So, an electric field was raised within the insulator due to this charge imbalance and it was practically negligible in the superconductor, that is, even for a short time the material behaved like a capacitor.   

The capacitors of Mahood~\cite{Mahood} are the unique devices in the table \ref{tab1} that are not made by rounded plates. In fact, multilayers capacitors were used in the experiments. We considered in this paper the first one of his experiments, in which was applied a 11 kHz sinusoidal frequency AC voltage on the capacitors. Our theoretical result shows that the weight variation is in agreement with the measured value one if we use the parameters described in his article. 

In the table \ref{tab1}, the last line corresponds to our experimental result for the maximum loss weight obtained, as measured by a digital scale. The experimental setup for the symmetrical capacitor sample used can be analyzed in detail in Fig.~1. It is important to say that when one maintains constant the voltage, independently of the distance between the capacitor and the scale, the measurements were always the same, a feature that eliminates the ionic wind hypothesis due to the independence of the distance. Besides, as an additional checking, we implemented measurements in which the capacitor was plated with symmetry axis parallel to the tray of the scale. We obtained a null result up to measurements around 14 kV, so that one concludes that there is not electrostatic effect between the components. However, it is important to comment that measurements with that scale presented high difficulty in the reading of the display, because of the many fluctuations on the values. In that first setup, we made a lot of measurements on one capacitor sample, so that we could read in the scale the higher weight loss for the maximum value of tension provided by the power supply. Despite of the accuracy in the value measured, the possible interference of seismic and thermal effects, the dependence on the humidity and the critical influence of the insulating materials guided us to improve the experimental assembly. Hence, it was needed to implement an scheme for a more stable reading with a more sophisticated measurement device. In relation to our improved experimental setup, we constructed five symmetrical capacitors and performed some measurements of weight variation of them by considering the scheme shown in Fig.~2. 
\begin{figure}[h]
\begin{center}
\includegraphics[width=8cm]{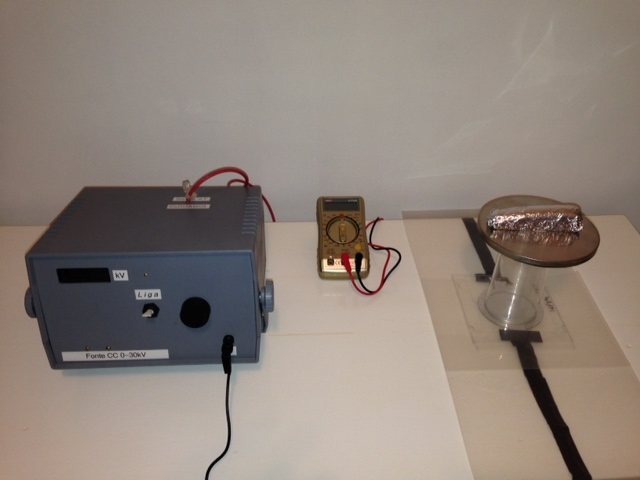}
\caption[dummy0]{Scheme of the experimental setup used for the measurements of the weight losses (in mgf) of the same symmetrical capacitor earlier described as function of the voltage applied (in kV). The setup represents an advance in relation to avoid noise interferences in the data acquisition due to the implementation of an accelerometer, that allows us to obtain three thousand and forty samples per second of input data for each voltage applied in its  maximum resolution. The distance between the accelerometer and the power supply was increased and all the devices remained in the same plane on the work table in order to reduce possible ionic wind effects. Besides, all the connections were insulated by layers of polypropylene. The z-axis of the accelerometer sensor was placed in the upward vertical direction and parallel to the symmetry axis of the capacitor. A 12 cm high cylinder of glass supported a plastic tray covered by a shielding aluminium where it was enclosed the accelerometer. } 
\label{fig2}
\end{center}
\end{figure}
\vskip 1cm
\begin{figure}[hb]
\begin{center}
\includegraphics[width=9.2cm]{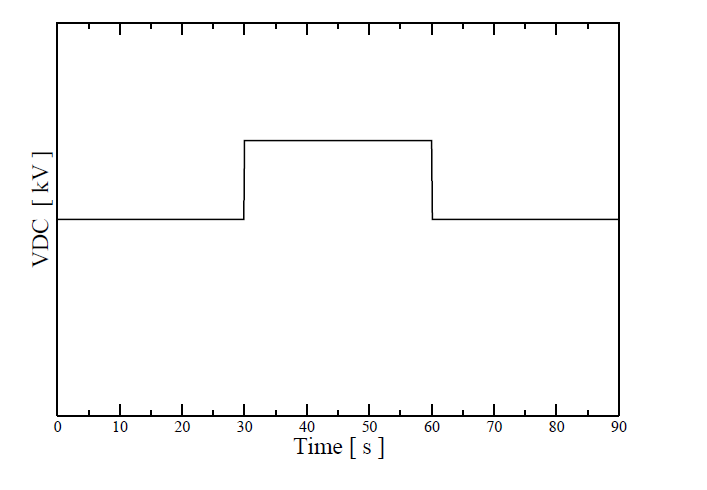}
\caption[dummy0]{Scheme of the experimental procedure adopted when using the high resolution accelerometer. One can realize that the procedure is executed in two steps or phases. In the first one, the DC voltage was null and the accelerometer was measuring the gravity acceleration since the beginning ( t = 0s when it was turned on ). The second phase corresponds to the period between 30 s and 60 s, in which the power supply was turned on.} 
\label{fig3}
\end{center}
\end{figure}

So, in that second step of the experimental work, we implemented a significant improvement in the scheme of the measurements, adopting a accelerometer as a weight variation sensor. The main advantages with the implementation of such a scheme based on the accelerometer were to obtain automatically hundreds of measurements for each value of tension, so that thermal, electromagnetic, acoustic and seismic interferences could be realized in a clear way and the standard deviations could be determined. We applied that experimental setup on our sample capacitors and slowly varied the tension. We measured the average weight reduction in mg (or mgf) for some values of the voltage applied on the capacitor up to 7 kV.
 
The analysis of the accelerometer measures were made in two periods of time after it is switched on; the first one was from 1th to 30th second and the following one was from 31th to 60th second. The high-voltage power supply was turned on between 30 and 60 seconds after switched on the accelerometer (see scheme in Fig.~3). Because of such a procedure, the first step of the analysis of accelerometer measures was made when the  voltage was switched off and the second step  was made when the voltage was switched on. This procedure was repeated for each different DC voltage applied, that is, from 1kV up to 7kV. There was no effect below 3kV. The anomalous force could be detected for voltages higher than 3kV, but interferences on the operation do not allow a positive conclusion on the presence of the effect for all of the operation time ranges. However, for the higher voltage around 7kV, we concluded that the anomalous force really arises, as one can see in the average values in the table \ref{averages}. The difference between two periods of average measures indicated an upward force even changing the polarity of the capacitor. It was considered in our analysis the difference in standard deviation between the "on" and "off" periods. The accelerometer was configured to provide high resolution sampling, measuring 320 samples per second. The cause of the anomalous force seems to be not related to the piezoelectric effect because the effect is proportional to the voltage applied and the measured force seems to be proportional to the squared voltage. Other reason is that the piezoelectric material contracts when the voltage is reverted and this effect was not observed in that experiment, considering that the dielectric material used in that capacitor was non-polar (piezoelectric materials are usually polar materials). By following that procedure, we obtain measurements that are consistent with the weight loss hypothesis, mainly for the higher voltage applied. Despite of the weakness of the effect, in hypothesis it could be enhanced if we had used a device with high physical values, as supercapacitance~\cite{Markoulidis} or by capacitances optically controlled through a dielectric constant~\cite{Yamasaki}.  

Our experiments also showed that the weight losses (in mg or in mgf) increased for higher voltage applied (in kV) to the plates of the symmetrical capacitors. It is worth to emphasize that the effects occurs in mgf scale and the theory is very consistent with such a order of magnitude for the weight losses. The theoretical values based on the theoretical framework were very close to the weight losses experimentally measured, mainly considering 
the standard deviations shown in the last column of table \ref{averages}, for the time of operation considered in the measurements. 
\begin{table}[h]
\caption{Measurements of average weight loss (in mgf) for 7 kV of the voltage applied (kV) on the sample capacitor, by considering long time intervals of collected data. The first column indicates the time in which the accelerometer collected data for the power supply turned off. The second one corresponds to the time of operation of the power supply. The third one shows for pair of time intervals considered the average weight loss measured by the accelerometer. In the last column, one can see the correspondent standard deviations calculated.}
\begin{tabular}{|c|c|c|c|}
\hline
Interval 'Off' (s) & Interval 'On' (s) & Weight Loss (mgf) & Error (mgf)\\
\hline 
1-30 &  31-60 & 36.8125 & 4.7 \\
1-30 &  35-60 & 36.7344 & 3.6 \\
1-30 &  40-60 & 36.7031 & 2.4 \\  
\hline
\end{tabular}
\label{averages}
\end{table}  

In the Fig.~4, we show the behavior of the measurements realized with the high-resolution accelerometer for the voltage 7 kV, by adopting the experimental procedure described in this work. The deadband setting is expressed in g / counts units and is applied to the output of z-axis. For the maximum resolution adopted in our experiment, the number of countings or the deadband value can be set to the integer 16384. So, the reading in the z-axis provides the local value of g if we divide the results by 16384 and the weight loss if we multiply by the mass of the capacitor setup. Hence, the differences between the measurements in the "off" and "on" phases showed a good precision in the order of mgf, with nearly constant value for many different time intervals analyzed. 

\vskip .75cm
\begin{figure}[h]
\begin{center}
\includegraphics[width=10.8cm]{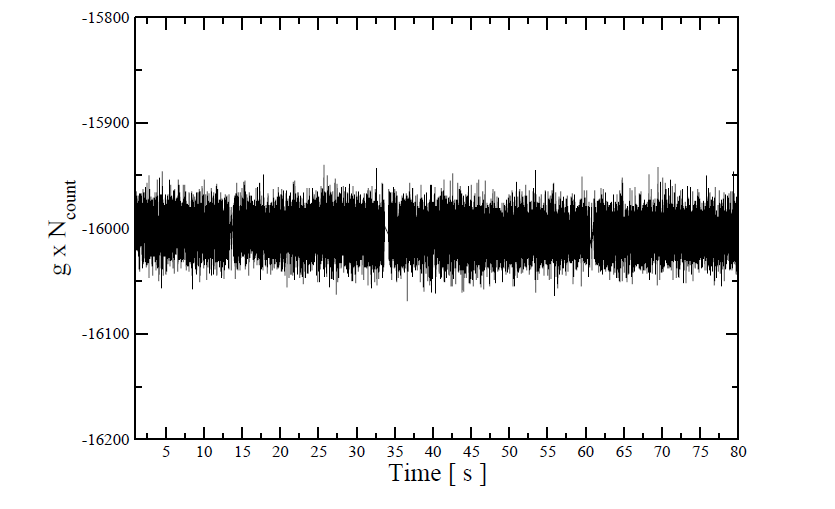}
\caption[dummy0]{Plot of the weight measurements of the symmetrical capacitor device ( in units of g / count, in which g is the standard gravity acceleration ) as a function of the operation time ( in s ) of the accelerometer, according to the apparatus shown in Fig.~2.} 
\label{fig4}
\end{center}
\end{figure}

We now present in the next section the theoretical proposal in order to explain the effect. Provided by such a theoretical proposal, we could explain our experimental results and the experiments realized by other authors, as we show in the next section. 
    
\section{Theoretical Description}
\label{sec:3}

In the 19th century, the relationship between macroscopic observables like electric susceptibility $\chi_e$ and individual properties of atomic or molecular electric dipoles such as polarizability $\alpha$ was firstly realized by Mossotti and Clausius for nonpolar materials~\cite{CM}. 

The model proposed in this work is based on the theoretical description of a set of the electric dipoles, subjected to a high electric potential. By means of that proposal, we calculated the magnitude of the forces generated on the symmetrical capacitors by employing the Clausius-Mossotti equation:
\begin{equation} 
\frac{\epsilon - 1}{\epsilon + 2} = \frac{4\pi N \alpha}{3}, 
\label{eqcm} 
\end{equation} 
in which $N$ is the number of particles per volume and $\alpha$ is the polarizability of the atom or molecule. Basically, Clausius-Mossotti equation presents in the left side the macroscopic variables and in the right one the microscopic variables. That equation provides the density of dielectric dipoles in a dielectric medium, based on its relative permittivity. The utilization of the equation is worth for solids and for low dielectric constant. 

The electric dipolar force F can be calculated according to Clausius-Mossotti relation:
\begin{equation}
F =  \frac{ 0.102 } {16 . \pi^2 }  \frac{\epsilon_r -1}{\epsilon_r + 2} \ \epsilon_0  A  E^2 , 
\label{cm1}
\end{equation}
in which the force $F$ is in units of Kgf, $A$ is the area of the capacitor plate, $\epsilon_r$ is the relative permittivity of the material, $\epsilon_0$ is the dielectric constant of the vacuum and $E$ is the electric field applied in the medium. 

The relative permittivity $\epsilon_r$ of the materials can be applied directly in the Clausius-Mossotti relation used in the dipolar force $F$ formula for nonpolar materials such as glass and wax, as referred in real experiments mentioned in this work. In case of plastics (polymers), the electronic displacement is reduced by the long chains of molecules when a weak electric field is applied. This is the case of our experiments where 1mm polystyrene thin layer was used as a capacitor dielectric layer. But even in that case the Clausius-Mossotti formalism can help us to get the right calculation.

We started the calculation of the electric susceptibility $\chi_e$ for the capacitor dielectric material, that is, the monomer C$_8$H$_8$, considering that $\chi_e = \epsilon_r - 1$, using the Clausius-Mossotti formalism:
\begin{equation}
\epsilon_r - 1 = \frac{4 \pi \rho \alpha}{m}, 
\label{hidro}
\end{equation}
in which $\rho$ is the density of the material and $m$ is the mass of the atom or molecule.
The polarizability is approximately the atomic of molecular radius cube. The result of this calculation is $\epsilon_r - 1 = $ 0.2378 for C$_8$H$_8$ parameters such as $\rho$  = 910 Kg/m$^3$, $m$ = 1.726 $\times$  10$^{-25}$ Kg, $\alpha$  = 3.59 $\times$ 10$^{-30}$ m$^3$. This result was used for the dipolar force $F$ calculation considering 7kV of  voltage, but the value is higher than the experimental measurements. The main reason of this remarkable difference is that the binding strength of C$_8$H$_8$ monomers in the molecular polystyrene structure is strong and the Clausius-Mossotti formalism has higher validity for molecules and atoms diluted in the material.  The electronic displacement of C$_8$H$_8$ monomers is much more difficult when they comprise a large molecular chain such as the polystyrene. In order to improve our analysis, we calculated the specific electric susceptibility $\chi_e$ for the hydrogen atoms weakly bound  in the  styrene monomer (C$_8$H$_8$). The result of this calculation is $\epsilon_r - 1$ = 0.086 for the hydrogen parameters, that is, $\rho$ = 76.24 Kg/m$^3$, $m$ = 1.66 $\times$ 10$^{-27}$ Kg and $\alpha$~=~53 $\times$ 10$^{-12}$ m$^3$. This result was used for the dipolar force $F$ calculation considering 7kV as voltage applied. 

The equation of dipolar force~(\ref{cm1}) can explain the effects measured with the experimental setups cited in the table 1. In the cases in which we have nonpolar materials, we can directly apply the equation~(\ref{cm1}) with no correction. For the material used in our experiments, we can calculate the dipolar forces by considering hydrogen atoms weakly bound in the styrene monomer (C$_8$H$_8$). 

The theoretical result of weight loss 40,38 mgf was also compared with our experimental measurements for the voltage 7kV, in the case of data acquisition with high resolution accelerometer, shown in the table 2. The values of that table show the dependence of the weight loss on the time operation of the accelerometer, for the "on" phase and the "off" phase, as described in the experimental description of this work. All of the values in the table 2 show a good stability ( average weight loss $\Delta {\overline{W}} \cong $ 30 - 37 mgf ) in the measurements of the average weight loss, despite of intense interference in the device. The average results are consistent with the hypothesis of existence of Biefeld-Brown effect, by considering long time intervals, so that possible interferences on the measurements have their effects reduced in the data acquisition.    

By considering our theoretical explanation, we calculated the values of weight losses measured in 
mgf~\cite{Manitoba,Mahood,Honda,Musha0} or anomalous force in N~\cite{Poher} in the experiments reported by those other authors and in our experiments, as one can see in the table 3. In the latter, we compare the measurements of weight losses of all the experiments reported in this work and the theoretical forecast based on the Clausius-Mossotti formalism. The theoretical calculations that we show in the table 3 consider the Eq.~(\ref{cm1}) to determine the magnitude of the weight loss in all of the cases, with exception of the cases involving styrene, in which the hydrogen hypothesis needs to be considered, that is, the force dipolar has to be calculated by means of the correction given by Eq.~(\ref{hidro}). We also show in the last line our result for the maximum voltage applied. 
\begin{table}[t]
\begin{centering}
\caption{Comparison between the several measurements of weight loss or anomalous force reported 
in Refs.~\cite{Manitoba,Mahood,Honda,Poher,Musha0} and the respective theoretical calculations based on the macroscopic observables. All the experimental setups labeled by the numbers in the first column involve direct measurements of weight losses, with exception of the work in Ref.~\cite{Poher}, in which the apparatus was mounted for directly measuring the raised force. The last line in the table indicates our 
experimental result for the maximum voltage applied, measured by the electronic scale. }
\begin{tabular}{|l|l|l|}
\hline
Experiment & Weight Loss or & Theoretical  \\
Setup & Force Measured & Forecast \\
\hline 
1~\cite{Manitoba}         	&380g	               &536g\\
1~\cite{Manitoba}	         	&30g	               &21.43g\\
2~\cite{Manitoba}		        &190g	               &268g\\
2~\cite{Manitoba}		          &20g	               &10.75g\\
3~\cite{Manitoba}		          &0.735 g	           &0.718g\\
4~\cite{Honda}		  &(0.9 $\pm$ 0.6)g	         &(0.922 $\pm$ 1.04)g\\
5~\cite{Poher}		        &7800N	             &7800N\\
6~\cite{Poher}		        &156N	               &156N\\
7~\cite{Poher}		      &3900N               &3900N\\
8~\cite{Mahood}		&5.10$^{-5}$ N	                       & 12.10$^{-5}$ N\\
9~\cite{Musha0} & 10.9 mg &	11.29mg  \\
9~\cite{Musha0} & 0.825 mg	& 0.71mg  \\
10~\cite{Musha0} & 0.6 g	& 0.9 g\\
11 & 220 mg	& 245 mg\\
\hline
\end{tabular}
\end{centering}
\label{results}
\end{table}  

Our experiments also showed that the weight losses (in mg or in mgf) increased for higher voltage applied (in kV) to the plates of the symmetrical capacitors. It is worth to emphasize that the effects occurs in mgf scale and the theory is very consistent with such a order of magnitude for the weight losses. 

It is remarkable to observe that the theoretical result for hydrogen are in good accordance with the experimental measurements. The binding strength of hydrogen atoms in the styrene monomer is four times lower than carbon atoms one and the hydrogen mass is twelve times lower than carbon. Considering this, the electronic displacement in the hydrogen atoms is much easier than the carbon atoms. This is a probable reason that the hydrogen electric susceptibility makes the results of the dipolar force $F$ more compatible with the experimental results. The theoretical and experimental results indicate that the net polarizability of C$_8$H$_8$ monomers that are bound in the polystyrene structure probably has the strong contribution of the hydrogen atoms. This specific electric susceptibility $\chi_e$ = 0.086 for the hydrogen atoms bounded in the styrene  monomer (C$_8$H$_8$) can be used for the dipolar force $F$ calculation in other experiment~\cite{Musha0}, in which a 0.2 mm thickness plastic dielectric material was used between the rounded plates of the capacitor with radius 0.0325 m. The theoretical results like 11.29mg for 1200VDC voltage applied and 0.71mg for 300VDC have good accordance with the respective experimental results as 10.9mg and 0.825mg.  

\section{Concluding Remarks} 
In this work, we present experimental and theoretical investigations concerning to the possibility of existence of an anomalous force on symmetrical capacitors, operating in high voltage, motivated by a lot of earlier experimental and theoretical works described in the literature from other different authors. Our experimental measurements indicated the presence of a small but persistent weight loss in the scale of mg, for different values of the high voltage applied on the symmetrical capacitors, with maximum value for the weight loss 220 mg, for the maximum voltage applied ( up to 20 kV ) on the device sample. In order to improve the accuracy of the measurements, due to high instabilities in some experimental data obtained from the digital scale reading, we have implemented an experimental method based on the utilization of a high resolution accelerometer, that collects hundreds of data per second. Our experimental results confirm the presence of a upward force acting on the system in both assemblies. We also propose an expression in the macroscopic level of a dipolar force based on the Clausius-Mossotti relation. We show that by means of such a theoretical proposal one can explain the manifestation of the effects caused by the microscopic dipoles under high voltage on the macroscopic medium. In fact, the theoretical results show that such a concept can explain with good precision most of the experimental data in the literature. In relation to our experimental measurements of weight losses of symmetrical capacitor samples with a high resolution accelerometer, the theoretical results of the anomalous upward dipolar force based on Clausius-Mossotti expression present good agreement. The magnitude of the anomalous force seems to be directly related to the dipolar force calculated according to the Clausius-Mossotti relation and its nature is deeply related to the collective action of the internal electric dipoles from the dielectrics of symmetric capacitors working in high voltages.

It would convenient also to report that we are elaborating other works based on a new experimental setup involving the detection of higher magnitudes of the anomalous forces and the effect of induction of forces at distance, having as source high voltage capacitors as other kind of devices.  

\ack

Elio B. Porcelli and Victo S. Filho thank to S\~ao Paulo Research Foundation (FAPESP) for partial support.

\end{document}